\documentclass[prd,nobibnotes,twocolumn,showpacs]{revtex4}

\usepackage{graphics,dcolumn}
\newcommand{\be}{\begin{equation}}
\newcommand{\ee}{\end{equation}}
\newcommand{\Qa}[2]{\mathrm{Q_1}\! \left(\frac{#1}{#2}\right)}
\newcommand{\Qb}[2]{\mathrm{Q_2}\! \left(\frac{#1}{#2}\right)}
\renewcommand{\P}{{\mathcal{P}}}
\newcommand{\Pt}{{\partial_t\P}}
\newcommand{\R}{{\mathcal{R}}}
\newcommand{\K}{{\kappa}}
\usepackage{amsmath}
\usepackage{amssymb}

\newcommand{\lambdah}{{\lambda}}\newcommand{\gh}{{g}}

\allowdisplaybreaks[1]

\begin{document}
\title{Constraints on matter from asymptotic safety}
\author{Roberto Percacci}
\email{percacci@sissa.it}
\author{Daniele Perini}
\email{perini@he.sissa.it}
\affiliation{SISSA, via Beirut 4, I-34014 Trieste, Italy}
\affiliation{INFN, Sezione di Trieste, Italy}


\begin{abstract}
Recent studies of the ultraviolet behaviour of pure gravity
suggest that it admits a non-Gaussian attractive fixed point,
and therefore that the theory is asymptotically safe.
We consider the effect on this fixed point of massless 
minimally coupled matter fields. 
The existence of a UV attractive fixed point
puts bounds on the type and number of such fields.
\end{abstract} \maketitle

There is growing evidence that four dimensional Quantum Gravity may be
``asymptotically safe'' \cite{Weinberg:1979}, \emph{i.e.}\ that it
admits a UV stable fixed point (FP) of the renormalisation group (RG)
equations with a finite number of attractive directions. In such a
theory, the locus of points that flow towards the FP is a
finite-dimensional hypersurface called the ``critical surface''. One
can take the UV limit in a controlled way starting from any point
belonging to the critical surface. The coordinates on this surface
(the relevant couplings) are free parameters, to be determined
experimentally; the remaining ones are fixed. Since there is only a
finite number of arbitrary parameters, the theory is predictive.
Perturbative renormalisabity, along with asymptotic freedom,
corresponds to a special case of asymptotic safety in which the UV
limit is taken at a Gaussian FP (GFP).  Therefore, asymptotic safety
is a generalization of the standard notion of renormalisability in a
nonperturbative context.

When this proposal was first put forward, some
suggestive calculations were performed in $2+\epsilon$ dimensions
\cite{Weinberg:1979,Gastmans:1978ad}, 
but no one knew how to continue to $\epsilon=2$, which is the
physically interesting case, and the programme came to a halt for lack
of nonperturbative information. It now appears that a suitable tool
has been found to address this issue: it is an exact renormalisation
group equation, describing the change in the effective action as
a certain infrared cutoff $k$ is varied
\cite{Wetterich:1993yh,Polchinski:1984gv}. 

The specific form of the exact RG equation that we shall use here
was written originally in \cite{Wetterich:1993yh} for a scalar field $\phi$.
The classical or bare action $S$ is modified by the addition of a purely quadratic term
$\Delta S_k(\phi)=\int dx \phi \R_k(z)\phi$,
whose only effect is to modify the propagator of the fields by adding to
$z=-\nabla^2$ a cutoff term $\R_k(z)$.
The modified propagator is denoted $\P_k(z)=z+\R_k(z)$.
(In the following we will often omit the subscripts $k$ for notational
simplicity.)
The function $\R_k$ is chosen so as to suppress the propagation 
of the field modes with momenta less than $k$.  
For example, we work with a smooth cutoff
\begin{equation}
\R_k(z)=\frac{2a\,ze^{-a\left(\frac{z}{k^2}\right)^b}}
{1-e^{-a\left(\frac{z}{k^2}\right)^b}},%
\end{equation}
depending on two parameters $a$ and $b$.
The modified inverse propagator $\P_k$
gives rise to a $k$-dependent effective action $\Gamma_k$ which
is equal to the bare action when $k$ is equal to the UV cutoff
and tends to the ordinary effective action when $k\to 0$.

The action $\Gamma_k$ satisfies the following exact RG equation
\cite{Wetterich:1993yh}:
\begin{equation}
\label{eq:erge}
\partial_t\Gamma_k={1\over 2} {\rm Tr}
\left({\delta^2 \Gamma_k\over\delta\phi\delta\phi}+R_k\right)^{-1}
\partial_t R_k\ ,
\end{equation}
where $t=\ln k$. This is essentially an RG-improved one-loop equation.
In fact, the one-loop effective action in the presence of the cutoff
is given by 
$\Gamma_k^{(1)}=S+{1\over 2}\ln\det
{\delta^2(S+\Delta S_k)\over\delta\phi\delta\phi}$
which, upon taking a derivative with respect to $t$, yields
an equation identical to \eqref{eq:erge} except that $S$ appears in the r.h.s.\
instead of $\Gamma_k$.

This formalism was first applied to gravity (regarded as an 
effective field theory) in \cite{Reuter:1998cp}.
The resulting exact RG equation is very similar to \eqref{eq:erge},
except that all terms in the equation are matrices, and
one has to add the contributions of gauge-fixing and ghost terms.
In addition to the parameters $a$ and $b$, the gravitational effective
action $\Gamma_k$ also depends on a gauge-fixing parameter $\alpha$.
There are arguments to the effect that $\alpha$ tends to zero
in the UV limit \cite{Litim:1998nf}.
Unless otherwise stated, in this paper all results will be given for 
$\alpha=0$, $a=1/2$ and $b=1$. The dependence on these parameters
will be briefly discussed in the conclusions.

This exact equation can be made manageable by approximating
(``truncating'') the functional form of the action.  In
\cite{Reuter:1998cp} the truncated action was assumed
to have the form of a Euclidean Einstein-Hilbert action:
\begin{equation}
\label{eq:action}
\K\int\,\mathrm{d}^4x\sqrt{g}\left(2\Lambda- R\right)\ ,
\end{equation}
where
$\Lambda$ is the (dimension-two) cosmological constant and $\K=1/16\pi
G$, where $G$ is Newton's constant.  
Inserting this ansatz for $\Gamma_k$, 
with $k$-dependent coupling constants, 
into the exact equation, one can
extract the beta functions:
\begin{equation}\label{eq:beta.def}
\left\{
\begin{aligned}
\partial_t(\K\Lambda)&={1\over 2{\rm Vol}}
\partial_t\Gamma_k\Bigg|_{R=0}\\
\partial_t \K&=
-{1\over{\rm Vol}}
{\partial\over\partial R}\partial_t\Gamma_k\Bigg|_{R=0},\\
\end{aligned}\right.
\end{equation}
where Vol=$\int d^4x\sqrt{g}$,
whence one gets $\partial_t\Lambda$ and $\partial_t G$.
In \cite{Dou:1998fg} some
technical improvements were made and the effect of minimally coupled
matter fields on the running of the gravitational couplings was
calculated. 

The main result of those papers was to write down the beta functions
for the cosmological constant and Newton's constant, and finding the
general behaviour of the couplings in different regimes.  It was later
recognized \cite{Souma:1999at,Lauscher:2001ya} that this truncated
gravitational RG equation admits, in addition to a Gaussian FP (GFP)
at $(\Lambda,G)=(0,0)$, also a UV-attractive non-Gaussian FP (NGFP)
for positive $\Lambda$ and $G$, thus raising new hopes for the
asymptotic-safety programme.  It is then paramount to prove that the
latter FP is not an artifact of the truncation, but is a genuine
feature of the theory.  Furthermore, in order to be relevant to the
real world, the NGFP must still exist when we add gauge and matter
fields.

The NGFP has so far passed several tests
 \cite{Lauscher:2001cq,Lauscher:2002mb}. 
It has been shown to exist, when $d=4$, for every shape of the cutoff
that has been tried, whereas in other dimensions it only exists for
certain forms of the cutoff. The properties of the NGFP are only very
weakly dependent on the cutoff scheme, which suggests that they have a
genuine physical meaning.  It is also remarkable that the value of
Newton's constant at the NGFP is always positive in $d=4$, a result
that is not {\it a priori} evident.  Most important, it was shown to
be stable under the addition of an \mbox{$R^2$-term} to the action,
whereas the GFP disappears under this perturbation
\cite{Lauscher:2002mb}.  These encouraging results warrant further
examination of the properties of this NGFP.

In this paper we consider the effect of minimally coupled, massless
matter fields.
The beta functions of the cosmological constant and Newton's constant
are modified by the presence of these fields, so this generalization 
has the potential to kill the NGFP.
One may hope that the NGFP is always present, but something more
interesting actually happens: 
within the Einstein-Hilbert truncation, the NGFP exists only for some
combinations of matter fields, so that its existence (together
with the conditions of attractivity and positivity of Newton's constant), 
can actually be used to put bounds on the type and number of fields.

We assume that in addition to the graviton there are $n_S$ real scalar
fields, $n_W$ Weyl fields, $n_M$ Maxwell fields, $n_{RS}$ (Majorana)
Rarita-Schwinger fields, all minimally coupled.  
We neglect all masses and interactions of the matter fields.
In the case of gauge fields, this can be justified by asymptotic freedom;
in the case of a single massive scalar we have verified 
that an attractive FP occurs for vanishing mass. 
For other interactions it is an approximation whose validity will have to
be tested in the future.

Using the heat-kernel expansion, which is valid for $k^2\gg R$, one can
express $\partial_t\Gamma_k$ in terms of integrals of the form:
\begin{equation}
Q_n[f]=\frac1{\Gamma(n)}\int_0^{+\infty}\mathrm{d}z\,z^{n-1}f(z).
\end{equation}
Then, using equations \eqref{eq:beta.def}, one obtains:
\begin{widetext} \begin{subequations}\label{eq:beta.functions}
\label{eq:beta.LAMBDA} \begin{align}
\partial_t(\K\Lambda)&= \frac{1}{64\pi^2} 
\Bigg\{
\left(n_S-2n_W+2n_M-4n_{RS}\right)\Qb{\Pt}{\P} 
+\Qb{\Pt(2\P+8\Lambda)}{\P\left(\P-2\Lambda\right)}
+\frac{\partial_t\K}{\K}
\Qb{\R\left(10\P-8\Lambda\right)}{\P\left(\P-2\Lambda\right)}
\Bigg\} %
\intertext{and} %
\label{eq:beta.G}
\partial_t\K&=\frac{1}{384\pi^2}
\Bigg\{
\left(-2n_S+4n_W-n_M\right)\Qa{\Pt}{\P}
+\left(-6n_W+9n_M-16n_{RS}\right)\,\Qb{\Pt}{\P^2}
\notag\\ 
&\hspace{12mm}
+\Qa{\Pt\left(13\P-10\Lambda\right)}{\P\left(\P-2\Lambda\right)}
+5\,\Qb{\Pt\left(11\P^2-12\P\Lambda+12\Lambda^2\right)}
{\P^2\left(\P-2\Lambda\right)^2}
\notag\\ &
+\frac{\partial_t\K}{\K}
\Bigg[
\Qa{\R\left(3\P+10\Lambda\right)}{\P\left(\P-2\Lambda\right)}
+5\,\Qb{\R\left(5\P^2+12\P\Lambda-12\Lambda^2\right)}{\left(\P-2\Lambda\right)^2}
\Bigg]
\Bigg\} .
\end{align}
\end{subequations} 
\end{widetext}
%
This agrees with \cite{Lauscher:2001ya} in the
absence of matter fields and with \cite{Dou:1998fg} upon expanding to
first order in $\Lambda$ \footnote{there is a coefficient $5/24$ in equations
(4.4) and (4.7) in \cite{Dou:1998fg} that should be replaced by
$-55/24$ and a coefficient $77/24$ in (4.9) that should read $17/24$.}.

Defining the dimensionless couplings $\lambda=\Lambda/k^2$ and 
$g=k^2G$, and their beta functions
$\beta_\lambda(\lambdah,\gh)=\partial_t\lambda=(\partial_t\Lambda/k^2)-2\lambdah$,
$\beta_g(\lambdah,\gh)\equiv\partial_t g=k^2\partial_t G+2g$, the
fixed points $(\lambda_*,\gh_*)$ are defined as the solutions of the
equations:
\begin{subequations}\label{system}
\begin{align}
\label{eq:system.a}\beta_\lambda(\lambdah,\gh)&=0\\
\label{eq:system.b}\beta_g(\lambdah,\gh)&=0\,.
\end{align}
\end{subequations}
Note that expressions \eqref{eq:beta.functions} contain the
derivative of $\kappa$ in the right-hand side, and therefore
in writing \eqref{system} one has to solve a linear system of two 
equations.
For the purpose of finding the fixed points, we will 
simplify the calculations by using the relation
$\partial_t\kappa/\kappa=-\partial_t G/G=2$,
which holds true at the fixed point, 
in the r.h.s.'s of \eqref{eq:beta.functions}.
Therefore, \eqref{eq:system.a} can be replaced by: 
\begin{equation}
\gh \cdot c (\lambdah)- 2 \lambdah=0,\tag{\ref{eq:system.a}'}
\label{eq:linear.g}
\end{equation}
where $c(\lambdah) $ is obtained by formally replacing
$G$ with 1 and $\partial_t G$ with $-2$ in the expression for
$\partial_t\Lambda/k^4$.
When $c(\lambdah)\neq 0$, we can solve \eqref{eq:linear.g} for $\gh$
and substitute the result into \eqref{eq:system.b}. We shall denote:
\begin{equation}\label{eq:notation}
h(\lambdah)=\beta_{\gh}\left(\lambdah,\frac{2\lambdah}{c(\lambdah)}\right),
\end{equation}
so that the zeroes of $h$ correspond to the FP's.
When $c(\lambdah)=0$, equation \eqref{eq:linear.g} implies
$\lambdah$=0. Therefore, if $c(0)\neq 0$ the only solution with
$\lambdah_*=0$ is the GFP, but if $c(0)=0$ we can have a NGFP with
$\lambdah_*=0$. Explicitly:
\begin{equation}
c(0)=\frac1{4\pi k^4}\!\left[\left(n_b-n_f\right)\!
\Qb{\Pt}{\P}\!+\!20\,\Qb{\R}{\P}\right],
\end{equation}
where $n_f=2n_W+4n_{RS}$ and $n_b=n_S+2n_M+2$ are the total numbers of
fermionic and bosonic degrees of freedom, so that the condition for
the existence of a NGFP with zero cosmological constant can be
restated as
\begin{equation}
\label{eq:bound}
\Delta=-\sigma ,
\end{equation}
where $\Delta=n_b-n_f$ and 
$\sigma= \!20\,\Qb{\R}{\P}/\Qb{\Pt}{\P}=5/\zeta(3)\approx 4.16$.
Due to the irrationality of $\sigma$, there is in general no combination
of matter fields that satisfies this condition. Still, the hyperplane
defined by \eqref{eq:bound} has an important physical significance: it
separates the regions with positive and negative $\lambdah_*$, as
shall become clear below.

Before discussing the general case, it will be useful to 
consider separately the effect of scalar and spinor fields. 
For all $n_S\geq 0$, the function $h$ vanishes at $\lambdah=0$,
corresponding to the GFP, and $h'(0)$ is always positive. 
Moreover, $h$ possesses two other zeroes, one positive and one negative.
The behaviour of $h$ for very large $n_S$ 
is given by:
\begin{equation} \label{eq:asymp.scal}
  h(\lambdah)\sim \frac{288 \pi \lambdah \left(\pi^2 \lambdah+18
      \zeta(3)\right)} {\left(\pi^2
    \lambdah+36 \zeta(3)\right)^2}\cdot \frac1{n_S}.
\end{equation}
which has a zero at $-\frac{18\zeta(3)}{\pi^2}\approx -2.19$
The negative zero of $h$ varies between $-\infty$ and $-2.19$ as 
$n_S$ ranges from 0 to $\infty$.
This FP is mixed for $n_S\leq 8$ and attractive for $n_S\geq9$, 
however $\gh_*$ is always negative
(it tends to 0$^-$ for $n_S\to\infty$);
therefore, this FP is not physically interesting.

\begin{figure}[t]
    \center{\resizebox{.85\columnwidth}{!}{\includegraphics{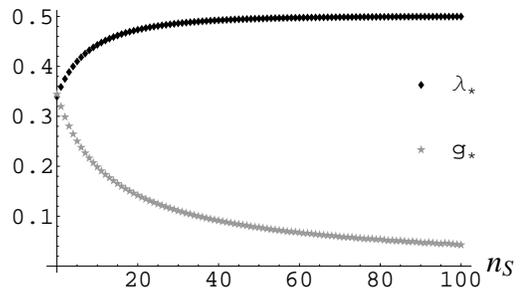}}
    \caption{\label{fig:1.1.2}$\lambdah_*$ and $\gh_*$ as functions of $n_S$.}}
\end{figure}

The positive zero occurs at $\lambdah_*=0.339$ (and $\gh_*=0.344$) for $n_S=0$
\cite{Lauscher:2001ya} and $\lambda_*$ tends to 0.5 for large $n_S$, implying
$\gh>0$.  From now on, we shall refer to this point as the NGFP.  Its
behaviour as a function of the number of scalar fields is shown in figure
\ref{fig:1.1.2}.  It exists for all values of $n_S$,
$\lambdah_\ast$ and $\gh_\ast$ always remain positive
and it is always attractive.

Fermions have the opposite effect: as $n_W$ grows, the cosmological
constant decreases, whereas Newton's constant increases. Some exact
results are shown in the following table:
\begin{center}
\setlength{\tabcolsep}{2mm}
\begin{tabular}{|c|D{.}{.}{3}|D{.}{.}{3}|}
\hline
$n_W$ & \multicolumn{1}{c|}{$\lambdah_*$} &  \multicolumn{1}{c|}{$\gh_*$} \\
\hline\hline
1 &  0.292 & 0.412 \\
\hline
2 &  0.211 & 0.524 \\
\hline
3 &  0.027 & 0.769 \\
\hline
4 & -1.379 & 2.295 \\
\hline
5 & -3.143 & 3.340 \\
\hline
6 & -4.337 & 3.673 \\
\hline
8 & -6.180 & 3.841 \\
\hline
10 & -7.640 & 3.794 \\
\hline
20 & -12.328 & 3.106 \\
\hline
50 & -17.896 & 1.839 \\
\hline
100 & -20.858 & 1.080 \\
\hline
\end{tabular}
\end{center}

The asymptotic behaviour for
large $n_W$ is:
\begin{equation}
  \label{eq:asymp.ferm}
  h(\lambdah)\sim\frac{-144\pi \lambdah\left[ \left( {\pi }^2 -9 \right)\lambdah + 18 \zeta(3) \right] }
  {{\left[ \left(\pi^2 -9 \right) \lambdah + 36\zeta(3)
      \right]}^2}\cdot {1\over n_W},
\end{equation}
which has a zero at $\lambdah=-\frac{18\,Z(3)}{{\pi }^2-9}\approx
-24.9$: this is the point that is approached asymptotically by the
NGFP; in this case, too, $\gh\to 0^{+}$. The FP is attractive for all
values of $n_W$. For $n_W$ large, $\lambda_*$ becomes
quite large in absolute value, so when one goes on shell the
conditions for the validity of the heat-kernel expansion are violated.

The result up to this point is that the existence and attractivity
of the NGFP are not affected by the addition of scalar or Weyl fields
separately. When both types of fields are present at the same time,
however, restrictions begin to appear.

The function $h(\lambda)$ always has a zero at the origin, corresponding
to the GFP.
The sign of the derivative of $h(\lambda)$ at the origin is determined
by the parameter $\Delta'=\Delta+\sigma$: it is positive (negative) when $\Delta'$
is positive (negative).
Since $h(\lambdah)$ tends to $-\infty$ for $\lambda$ somewhere
between 0 and 1/2 (namely where $c(\lambda)=0$),
when $\Delta'>0$ there always exists a NGFP with positive $\lambdah_*$.
On the other hand, when $\Delta'<0$ $h$ has no positive zeroes and 
the existence of the NGFP for negative $\lambdah_*$ hinges on the 
asymptotic behaviour of $h$ for $\lambdah\to -\infty$: 
it only exists if $h$ tends to a negative asymptote.
The asymptotic behaviour of $h$ is given by 
$\lim\limits_{\lambda\to -\infty}h(\lambda)=192\pi/\tau$,
where
\begin{equation}
\label{eq:limith}
\begin{aligned}
\tau=&\tau_0+n_S\tau_S+n_W\tau_W+n_M\tau_M+n_{RS}\tau_{RS}\\
\tau_0=&-5\Qa{\Pt}{\P}-15\Qb{\Pt}{\P^2}+10\Qa{\R}{\P}
\\&+30\Qb{\R}{\P^2}\approx -21.45\\
\tau_S=&2\Qa{\Pt}{\P}\approx 6.580\\
\tau_W=&-4\Qa{\Pt}{\P}+6\Qb{\Pt}{\P^2}\approx -1.159\\
\tau_M=&\Qa{\Pt}{\P}-9\Qb{\Pt}{\P^2}\approx -14.71\\
\tau_{RS}=&16\Qb{\Pt}{\P^2}\approx 32.00
\end{aligned}
\end{equation}

The space spanned by the variables $n_S$, $n_W$, $n_M$ and $n_{RS}$
can thus be divided into four regions, labelled I, II, III and IV, 
depending on the sign of the two parameters $\Delta'$ 
and $\tau$.

Region I ($\Delta'>0$ and $\tau<0$) admits a single NGFP
with positive $\lambdah_*$
which turns out to be always attractive; pure gravity falls in this category. 
Region II ($\Delta'>0$ and $\tau>0$) still has a positive-$\lambdah_*$ 
NGFP, 
but also a negative one. The latter FP yields negative $g_*$, 
as discussed above in the case of pure scalars, and is physically
uninteresting;
the former is attractive, except for a wedge-shaped area adjacent to the
plane $n_f=n_b$.
In the region that we have explored, for $n_S<500$, $n_W<250$, $n_M<50$,
this area can be described by the equations $\Delta'>0$ and 
\begin{equation}
n_W>0.45 n_S+1.12 n_M +2.6\ ,
\end{equation}
with an error of order one in each variable.
Within this region, the NGFP appears to be either repulsive or mixed;
the numerical calculations are not to be trusted, however.
In principle, mixed points are not incompatible
with asymptotic safety, but they would require that only a specific
one-dimensional trajectory in the $\Lambda$-$G$ plane is physically
admissible.
Region III ($\Delta'<0$ and $\tau<0$) has a single attractive NGFP with 
negative $\lambda_*$ and positive $g_*$.
Finally in region IV ($\Delta'<0$ and $\tau>0$) there is no NGFP.

The value of $\lambda_*$ in regions I and II is always less that 1/2 
(and much smaller for larger values of the parameter $a$) and therefore
reasonably within the bounds of the approximations. On the other hand 
in region III $\lambda_*$ becomes quickly rather large in absolute value;
in this regime $R\gg k^2$ on shell and therefore the heat kernel
approximation ceases to be valid. 
In this region the results are only reliable close to the surface 
$\Delta'=0$.

The gray area in figure 2 shows the region of existence of the NGFP
in the presence of both scalar and Weyl fields. 
The additional lines give the boundaries of the permitted region for
$n_{RS}=0$ and $n_M=10,20$ (growing upwards and rightwards). 
We see that the permitted region becomes larger with increasing $n_W$.
On the other hand, it appears from (\ref{eq:limith}) that Rarita-Schwinger
fields have opposite effect of the Maxwell fields and therefore reduce
the size of the permitted region in the $n_S$-$n_W$ plane.

\begin{figure}[!t]\vspace*{3mm}
\centering{\resizebox{.85\columnwidth}{!}{\includegraphics{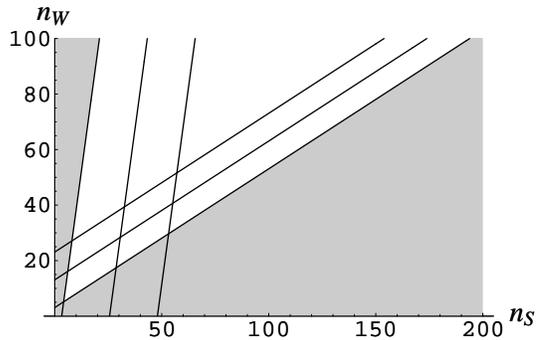}}}
\caption{\label{fig:3D.Nm.0}The UV attractive region (regions I, II and III) 
for $n_M=n_{RS}=0$ (in gray). Region IV is white. 
The other lines correspond to $\Delta'=0$ or $\tau=0$ 
for $n_M=10,20$.}\end{figure}

Many popular unified theories lie within the permitted region.
For example the standard model and the (now ruled out) minimal SU(5) 
GUT  with three generations lie in region III.
The SO(10) model lies in region I or III depending on the
pattern of symmetry breaking.
Supersymmetric theories have equal numbers of bosonic and fermionic
fields and therefore lie close to the surface
$\Delta=-\sigma$ (just below one of the diagonal lines in Fig.2).
If they have too few gauge fields (compared to the number
of scalars and fermions) they lie in the wedge of non-attractivity 
of region II while those with sufficiently many gauge fields
lie near the boundary between regions I and III
and therefore have an attractive NGFP.
The popular minimal supersymetric SU(5) model
falls into the first category.

Physical results are independent of gauge and cutoff parameters in the
exact theory, so the extent of parameter dependence that is observed
in the truncated theory gives a quantitative measure of the errors.
We have performed various tests on the parameter dependence of our
results and it is reassuring for the reliability of the truncation
that this dependence turns out to be reasonably mild.  For example,
the constant
$\sigma$ in (\ref{eq:bound}) is independent of the gauge parameter and
varies from 4.745 for $a=0.05$ to 2.765 for $a=20$. 
There are also other choices of the cutoff procedure that would yield 
the natural value $\sigma=0$.
The parameter $\tau$ is also gauge independent (except for a discontinuity
at $\alpha=0$; we have given here the values calculated for $\alpha=0$
rather than the limits for $\alpha\to 0$).
The plane $\tau=0$ is shifted and also slightly 
rotated to the right as $a$ grows.
Thus, the allowed region becomes larger as $a$ grows.
We have decided to present the more restrictive bounds for $a=1/2$
because one can be more confident that points within this region are
truly attractive NGFP's. On the other hand, points that nearly miss
the bounds could still turn out to be attractive NGFP's in a more
accurate analysis. 
More details on parameter dependence will be reported elsewhere.  
Of course, we are also neglecting all effects due
to matter self-interactions; besides a shift in the position of
$(\lambda_*,g_*)$, the true NGFP may have nontrivial components also
along other directions in the space of all couplings. For our results
to be relevant to the real world, such additional couplings should be
small.

In conclusion, the attractivity of the NGFP in the \mbox{$\Lambda$-$G$} 
plane puts bounds on the matter content of a realistic theory of the world. 
It will be interesting to see how these bounds are modified by matter 
interactions.

\begin{acknowledgments}
We would like to thank S. Bertolini and M. Reuter for helpful discussions.
\end{acknowledgments}

\bibliographystyle{apsrev}
\bibliography{database}

\begin{thebibliography}{19}
\expandafter\ifx\csname natexlab\endcsname\relax\def\natexlab#1{#1}\fi
\expandafter\ifx\csname bibnamefont\endcsname\relax
  \def\bibnamefont#1{#1}\fi
\expandafter\ifx\csname bibfnamefont\endcsname\relax
  \def\bibfnamefont#1{#1}\fi
\expandafter\ifx\csname citenamefont\endcsname\relax
  \def\citenamefont#1{#1}\fi
\expandafter\ifx\csname url\endcsname\relax
  \def\url#1{\texttt{#1}}\fi
\expandafter\ifx\csname urlprefix\endcsname\relax\def\urlprefix{URL }\fi
\providecommand{\bibinfo}[2]{#2}
\providecommand{\eprint}[2][]{\url{#2}}

\bibitem[{\citenamefont{Weinberg}(1979)}]{Weinberg:1979}
\bibinfo{author}{\bibfnamefont{S.}~\bibnamefont{Weinberg}}, in
  \emph{\bibinfo{booktitle}{General Relativiy: An Einstein centenary survey}},
  edited by \bibinfo{editor}{\bibfnamefont{S.~W.} \bibnamefont{Hawking}}
  \bibnamefont{and} \bibinfo{editor}{\bibfnamefont{W.}~\bibnamefont{Israel}}
  (\bibinfo{publisher}{Cambridge University Press}, \bibinfo{year}{1979}),
  chap.~\bibinfo{chapter}{16}, pp. \bibinfo{pages}{790--831}.

\bibitem[{\citenamefont{Gastmans et~al.}(1978)\citenamefont{Gastmans, Kallosh,
  and Truffin}}]{Gastmans:1978ad}
\bibinfo{author}{\bibfnamefont{R.}~\bibnamefont{Gastmans}},
  \bibinfo{author}{\bibfnamefont{R.}~\bibnamefont{Kallosh}}, \bibnamefont{and}
  \bibinfo{author}{\bibfnamefont{C.}~\bibnamefont{Truffin}},
  \bibinfo{journal}{Nucl. Phys.} \textbf{\bibinfo{volume}{B133}},
  \bibinfo{pages}{417} (\bibinfo{year}{1978}).

\bibinfo{author}{\bibfnamefont{S.~M.} \bibnamefont{Christensen}}
  \bibnamefont{and} \bibinfo{author}{\bibfnamefont{M.~J.} \bibnamefont{Duff}},
  \bibinfo{journal}{Phys. Lett.} \textbf{\bibinfo{volume}{B79}},
  \bibinfo{pages}{213} (\bibinfo{year}{1978}).

\bibinfo{author}{\bibfnamefont{H.}~\bibnamefont{Kawai}} \bibnamefont{and}
  \bibinfo{author}{\bibfnamefont{M.}~\bibnamefont{Ninomiya}},
  \bibinfo{journal}{Nucl. Phys.} \textbf{\bibinfo{volume}{B336}},
  \bibinfo{pages}{115} (\bibinfo{year}{1990}).

\bibinfo{author}{\bibfnamefont{I.}~\bibnamefont{Jack}} \bibnamefont{and}
  \bibinfo{author}{\bibfnamefont{D.~R.~T.} \bibnamefont{Jones}},
  \bibinfo{journal}{Nucl. Phys.} \textbf{\bibinfo{volume}{B358}},
  \bibinfo{pages}{695} (\bibinfo{year}{1991}).

\bibinfo{author}{\bibfnamefont{H.}~\bibnamefont{Kawai}},
  \bibinfo{author}{\bibfnamefont{Y.}~\bibnamefont{Kitazawa}}, \bibnamefont{and}
  \bibinfo{author}{\bibfnamefont{M.}~\bibnamefont{Ninomiya}},
  \bibinfo{journal}{Prog. Theor. Phys. Suppl.} \textbf{\bibinfo{volume}{114}},
  \bibinfo{pages}{149} (\bibinfo{year}{1993}).

\bibinfo{author}{\bibfnamefont{T.}~\bibnamefont{Aida}},
  \bibinfo{author}{\bibfnamefont{Y.}~\bibnamefont{Kitazawa}},
  \bibinfo{author}{\bibfnamefont{J.}~\bibnamefont{Nishimura}},
  \bibnamefont{and} \bibinfo{author}{\bibfnamefont{A.}~\bibnamefont{Tsuchiya}},
  \bibinfo{journal}{Nucl. Phys.} \textbf{\bibinfo{volume}{B444}},
  \bibinfo{pages}{353} (\bibinfo{year}{1995}),
  \eprint[http://arXiv.org/abs]{hep-th/9501056}.

\bibitem[{\citenamefont{Polchinski}(1984)}]{Polchinski:1984gv}
\bibinfo{author}{\bibfnamefont{J.}~\bibnamefont{Polchinski}},
  \bibinfo{journal}{Nucl. Phys.} \textbf{\bibinfo{volume}{B231}},
  \bibinfo{pages}{269} (\bibinfo{year}{1984}).

\bibinfo{author}{\bibfnamefont{C.}~\bibnamefont{Bagnuls}} \bibnamefont{and}
  \bibinfo{author}{\bibfnamefont{C.}~\bibnamefont{Bervillier}},
  \bibinfo{journal}{Phys. Rept.} \textbf{\bibinfo{volume}{348}},
  \bibinfo{pages}{91} (\bibinfo{year}{2001}),
  \eprint[http://arXiv.org/abs]{hep-th/0002034}.

\bibinfo{author}{\bibfnamefont{J.}~\bibnamefont{Berges}},
  \bibinfo{author}{\bibfnamefont{N.}~\bibnamefont{Tetradis}}, \bibnamefont{and}
  \bibinfo{author}{\bibfnamefont{C.}~\bibnamefont{Wetterich}},
  \bibinfo{journal}{Phys. Rept.} \textbf{\bibinfo{volume}{363}},
  \bibinfo{pages}{223} (\bibinfo{year}{2002}),
  \eprint[http://arXiv.org/abs]{hep-ph/0005122}.

\bibitem[{\citenamefont{Wetterich}(1993)}]{Wetterich:1993yh}
\bibinfo{author}{\bibfnamefont{C.}~\bibnamefont{Wetterich}},
  \bibinfo{journal}{Phys. Lett.} \textbf{\bibinfo{volume}{B301}},
  \bibinfo{pages}{90} (\bibinfo{year}{1993}).

\bibitem[{\citenamefont{Reuter}(1998)}]{Reuter:1998cp}
\bibinfo{author}{\bibfnamefont{M.}~\bibnamefont{Reuter}},
  \bibinfo{journal}{Phys. Rev.} \textbf{\bibinfo{volume}{D57}},
  \bibinfo{pages}{971} (\bibinfo{year}{1998}),
  \eprint[http://arXiv.org/abs]{hep-th/9605030}.

\bibitem[{\citenamefont{Litim and Pawlowski}(1998)}]{Litim:1998nf}
\bibinfo{author}{\bibfnamefont{D.~F.} \bibnamefont{Litim}} \bibnamefont{and}
  \bibinfo{author}{\bibfnamefont{J.~M.} \bibnamefont{Pawlowski}}
  (\bibinfo{year}{1998}), \eprint[http://arXiv.org/abs]{hep-th/9901063}.

\bibitem[{\citenamefont{Dou and Percacci}(1998)}]{Dou:1998fg}
\bibinfo{author}{\bibfnamefont{D.}~\bibnamefont{Dou}} \bibnamefont{and}
  \bibinfo{author}{\bibfnamefont{R.}~\bibnamefont{Percacci}},
  \bibinfo{journal}{Class. Quant. Grav.} \textbf{\bibinfo{volume}{15}},
  \bibinfo{pages}{3449} (\bibinfo{year}{1998}),
  \eprint[http://arXiv.org/abs]{hep-th/9707239}.

\bibitem[{\citenamefont{Souma}(1999)}]{Souma:1999at}
\bibinfo{author}{\bibfnamefont{W.}~\bibnamefont{Souma}},
  \bibinfo{journal}{Prog. Theor. Phys.} \textbf{\bibinfo{volume}{102}},
  \bibinfo{pages}{181} (\bibinfo{year}{1999}),
  \eprint[http://arXiv.org/abs]{hep-th/9907027}.

\bibitem[{\citenamefont{Lauscher and
  Reuter}(2002{\natexlab{d}})}]{Lauscher:2001ya}
\bibinfo{author}{\bibfnamefont{O.}~\bibnamefont{Lauscher}} \bibnamefont{and}
  \bibinfo{author}{\bibfnamefont{M.}~\bibnamefont{Reuter}},
  \bibinfo{journal}{Phys. Rev.} \textbf{\bibinfo{volume}{D65}},
  \bibinfo{pages}{025013} (\bibinfo{year}{2002}{\natexlab{d}}),
  \eprint[http://arXiv.org/abs]{hep-th/0108040}.

\bibitem[{\citenamefont{Lauscher and
  Reuter}(2002{\natexlab{a}})}]{Lauscher:2001cq}
\bibinfo{author}{\bibfnamefont{O.}~\bibnamefont{Lauscher}} \bibnamefont{and}
  \bibinfo{author}{\bibfnamefont{M.}~\bibnamefont{Reuter}},
  \bibinfo{journal}{Int. J. Mod. Phys.} \textbf{\bibinfo{volume}{A17}},
  \bibinfo{pages}{993} (\bibinfo{year}{2002}{\natexlab{a}}),
  \eprint[http://arXiv.org/abs]{hep-th/0112089}.

\bibinfo{author}{\bibfnamefont{O.}~\bibnamefont{Lauscher}} \bibnamefont{and}
  \bibinfo{author}{\bibfnamefont{M.}~\bibnamefont{Reuter}},
  \bibinfo{journal}{Class. Quant. Grav.} \textbf{\bibinfo{volume}{19}},
  \bibinfo{pages}{483} (\bibinfo{year}{2002}{\natexlab{b}}),
  \eprint{hep-th/0110021}.

\bibitem[{\citenamefont{Lauscher and
  Reuter}(2002{\natexlab{c}})}]{Lauscher:2002mb}
\bibinfo{author}{\bibfnamefont{O.}~\bibnamefont{Lauscher}} \bibnamefont{and}
  \bibinfo{author}{\bibfnamefont{M.}~\bibnamefont{Reuter}},
  \bibinfo{journal}{Phys. Rev.} \textbf{\bibinfo{volume}{D66}},
  \bibinfo{pages}{025026} (\bibinfo{year}{2002}{\natexlab{c}}).


\end{thebibliography}

\end{document}